\input{aipcheck}

\documentclass[
    ,final            
  ]
  {aipproc}

\layoutstyle{8x11single}

\begin{document}

\title{Gluon Saturation in QCD at High Energy:\\
 Beyond Leading Logarithms}


\author{Guillaume Beuf}{
  address={Department of Physics,
Brookhaven National Laboratory,\\
Upton, NY 11973, USA.\footnote{Now at: Departamento de F\'isica de Part\'iculas and IGFAE, Universidade de Santiago de
Compostela, E-15706 Santiago de Compostela, Spain.}}
}

\begin{abstract}
Progresses towards the calculation and the understanding of NLO/NLL contributions to Deep Inelastic Scattering at low $x$ with gluon saturation are being reviewed.
\end{abstract}

\maketitle


\section{Introduction}

Hadronic collisions in the high energy limit are probing with a better and better time resolution the quantum fluctuations in the wave-functions of the incoming particles, due to time dilation.
On the contrary, due to length contraction, the longitudinal resolution is getting worse and worse. In the Regge limit, the typical transverse scales related to a given scattering process are kept fixed while the high energy limit of the collision is taken, so that the transverse resolution stays constant. Hence, in the Regge limit, a growing number of fluctuations of the incoming particles are participating to scattering processes, but those located in an volume set by the transverse and longitudinal resolutions cannot be distinguished and act coherently. Coherent multiple scattering of quarks and gluons is thus an unavoidable ingredient in the physics of the Regge limit. Equivalently, the coherent clusters in the hadronic wave-functions are more efficiently described in terms of semi-classical gluon fields than of partonic Fock states.

The physics of the Regge limit is then beyond the reach of the improved parton model, with parton distributions satisfying the DGLAP and/or BFKL evolution equations, in which the hadronic wave-functions are assumed to be dilute, and scattering processes are assumed to occur via a single binary collision of partons.
The phenomenon of gluon saturation \cite{Gribov:1984tu,Mueller:1985wy} corresponds to that breakdown of the improved parton model picture via the onset of high density effects. Gluon saturation as well as the following dense regime are properly described in the Color Glass Condensate effective theory \cite{McLerran:1993ni,McLerran:1993ka,McLerran:1994vd} and related formalisms.

These frameworks have been successfully applied to the calculation of various observables, related for example to forward physics at RHIC and at the LHC, and to early stages of heavy ion collisions. But the most straightforward and best understood application remains the one to deep inelastic scattering (DIS) at low Bjorken $x$. There is an ongoing effort for a few years to update from leading order (LO) and leading logs (LL) to next-to-leading order (NLO) and next-to-leading logs (NLL) the theory of DIS with gluon saturation. The purpose of the present contribution is to review the main progresses in that direction. But first, let us recall the LO/LL formalism and results.

\section{Gluon saturation at LO/LL accuracy}

The $F_T$ and $F_L$ DIS structure functions are proportional to the total cross sections for the collision between the target and a transverse or longitudinal virtual photon. At low $x$ and at LO accuracy, these total cross sections satisfy the dipole factorization \cite{Nikolaev:1990ja}: they can be written as a convolution of the probability of splitting of the photon into a quark-antiquark dipole before reaching the target, known from QED, with the imaginary part of the dipole-target elastic scattering amplitude, which is a non-perturbative quantity.

However, logarithmically divergent soft gluons emissions by the dipole are part of the higher order corrections. The corresponding leading logs can be resummed via a renormalization group evolution of the dipole-target scattering amplitude with respect to the total energy of the collision or to the available rapidity range $Y\sim \log(1/x)$. This evolution in $Y$ is given by the B-JIMWLK equations \cite{Balitsky:1995ub,Jalilian-Marian:1997jx,Jalilian-Marian:1997gr,Jalilian-Marian:1997dw,Kovner:2000pt,Weigert:2000gi,Iancu:2000hn,Iancu:2001ad,Ferreiro:2001qy},
which, for the DIS case, can be safely approximated by the BK equation \cite{Balitsky:1995ub,Kovchegov:1999yj,Kovchegov:1999ua}, which is a nonlinear extension of the BFKL equation.

The nonlinearity of the BK equation encodes gluon saturation and restores the unitarity of the dipole-target scattering amplitude at fixed impact parameter.
The impact parameter dependence of the dipole-target amplitude is usually neglected, in order to disentangle the weak coupling dynamics associated with gluon saturation from the long-range strong coupling QCD dynamics. Going beyond that approximation remains the main challenge in gluon saturation theory \cite{Berger:2010sh} and will not be discussed further here. The imaginary part $N(r,Y)$ of the dipole-target amplitude then depends only on $Y$ and on the dipole size $r$. For physical solutions of the BK equation, $N(r,Y)\propto r^2$ in the dilute regime at small $r$ and $N(r,Y)\rightarrow 1$ in the dense regime at large $r$. At each $Y$, the transition between these two regimes is encoded by the saturation scale $Q_s(Y)$ and due to the BK equation, $Q_s(Y)$ grows with $Y$ and $N(r,Y)$ scales to a very good approximation with $r^2 {Q_s}^2(Y)$ after a few units in $Y$. This is in qualitative agreement with the geometric scaling property found in the HERA data \cite{Stasto:2000er}.

The BK equation belongs to a class of nonlinear equations whose solutions have a universal asymptotic behavior, independent of the initial condition \cite{bramson,vanS}. In particular, the large $Y$ behavior of $Q_s(Y)$ writes
\begin{equation}
\log {Q_s}^2(Y) = \lambda Y + c_1\, \log Y + \textrm{Const}. + c_2\, Y^{-1/2} + {\cal O}(Y^{-1})\, ,
\end{equation}
with $\lambda$, $c_1$ and $c_2$ three known universal constants \cite{Mueller:2002zm,Munier:2003sj,Munier:2004xu}.
However, for any realistic value of the coupling $\alpha_s$, the theoretical value $\lambda \simeq 4.88\, N_c\, \alpha_s/\pi$ is much larger than the value $\lambda \simeq 0.2$ or $0.3$ extracted from the HERA data.
This too fast evolution of $Q_s(Y)$ despite the qualitative successes of gluon saturation shows the need for NLO/NLL corrections for a quantitative description of gluons saturation physics.

\section{Running coupling and other NLO/NLL effects}

In order to reach the NLO/NLL accuracy, we need both the NLO contributions to the DIS cross section, and the NLO corrections to the BK equation so that it can resum high-energy next-to-leading logs.
Both types of corrections have been calculated recently, in Refs. \cite{Balitsky:2010ze} and \cite{Balitsky:2008zz,Balitsky:2009xg} respectively. However, more work is needed before a practical implementation of those results. Indeed, the NLO corrections to the DIS cross section are not yet available in the form generalizing the dipole factorization. And various large NLO corrections arising in the BK equation need to be resummed, in a similar way as for the BFKL equation \cite{Ciafaloni:1999yw,Ciafaloni:2003rd,Altarelli:2005ni}.
These large NLO corrections 
have three origins: the running of the coupling $\alpha_s$, the kinematical approximations usually performed to calculate the high-energy LL and NLL contributions, and the non-eikonal parts of the DGLAP splitting functions.

The large corrections of kinematical origin come from the incorrect ordering of successive gluons or dipoles in standard formalism. Leading logs indeed arise only when gluons are strictly ordered in $k^+$ and in $k^-$ simultaneously. Imposing this double ordering already in the LO equation then allows to resum large corrections of that type \cite{Motyka:2009gi}. A full treatment of these effects will be soon available \cite{BeufToAppear}. Concerning the non-eikonal corrections to the BK equation, only a crude treatment is available \cite{Gotsman:2004xb}, so that more work is needed.

The most studied and best understood of the large corrections are the ones related to the running of $\alpha_s$. A subset of the NLL contributions can be interpreted as LL contributions multiplied by the first correction in the perturbative expansion of the running coupling. These particular NLL contributions should not be included in the NLO corrections to the BK equation, but used instead to determine with what transverse scale $\alpha_s$ runs in the BK equation. However, there is an ambiguity in this separation of NLL contributions in two types, which leads to various prescriptions for the running of the coupling \cite{Balitsky:2006wa,Kovchegov:2006vj}. It is now widely accepted, both for theoretical and technical reasons, that Balitsky's prescription, proposed in Ref. \cite{Balitsky:2006wa}, is the most appropriate.

As an intermediate step between LO/LL and NLO/NLL accuracy, many studies have been dedicated to running coupling versions of the LO BK equation, in particular with Balitsky's prescription or more naive ones. In this context, a very good global fit of the HERA data has been performed \cite{Albacete:2010sy}. In that fit, however, $\Lambda_{QCD}$ is taken as a fit parameter and is found significantly smaller than the physical value. 

Analytical studies \cite{Mueller:2002zm,Munier:2003sj} have shown that when $\alpha_s$ runs, the asymptotic behavior of the solutions of the BK equation is still universal, but in a different universality class than in the fixed coupling case. In particular, $Q_s(Y)$ now behaves as \cite{Mueller:2002zm,Munier:2003sj,Beuf:2010aw}
\begin{equation}
\log ({Q_s}^2(Y)/\Lambda_{QCD}^2) = C_{1/2}\, \sqrt{Y} + C_{1/6}\, Y^{1/6}+ C_{0} + C_{-1/6} \, Y^{-1/6}+C_{-1/3} \, Y^{-1/3} + {\cal O}(Y^{-1/2})\, ,\label{QsRC}
\end{equation}
with $C_{1/2}$, $C_{1/6}$, $C_{0}$, $C_{-1/6}$ and $C_{-1/3}$ known and independent of the initial conditions. Moreover, $C_{1/2}$, $C_{1/6}$ and $C_{-1/6}$ are totally insensitive to possible NLO and higher order corrections to the BK equation, whereas $C_{0}$ and $C_{-1/3}$ depend on LO and NLO contributions, but not on NNLO or higher \cite{Beuf:2010aw}. Hence, at high enough $Y$, the main effect of NLO corrections to the BK equation apart from the running of $\alpha_s$ is a shift of $C_{0}$ which can be absorbed by a rescaling of $\Lambda_{QCD}$. This explains the need for such a rescaling in the fit of Ref.\cite{Albacete:2010sy}, where those NLO corrections are not included.

The asymptotic behavior \eqref{QsRC} is consistent with numerical simulations of the BFKL equation with running $\alpha_s$ at LO, strict NLO, or NLO with resummations, with toy model gluon saturation effects included \cite{Avsar:2011ds}. These numerical simulations also demonstrate the need to resum the large NLO corrections in order to obtain sensible NLO results, in particular at not too large $Y$.


\begin{theacknowledgments}
The works presented here have been performed under the Contract No. \#DE-AC02-98CH10886 with the
U.S. Department of Energy.
\end{theacknowledgments}



\bibliographystyle{aipproc}   

\bibliography{MaBiblioHEQCD}

\IfFileExists{\jobname.bbl}{}
 {\typeout{}
  \typeout{******************************************}
  \typeout{** Please run "bibtex \jobname" to optain}
  \typeout{** the bibliography and then re-run LaTeX}
  \typeout{** twice to fix the references!}
  \typeout{******************************************}
  \typeout{}
 }

\end{document}